%% file: Main.tex
\documentclass{ar-1col-S2O}
\usepackage[numbers]{natbib}
\usepackage{url}
\usepackage[utf8]{inputenc}
\usepackage{textgreek}
\usepackage{amsmath, amssymb}

\jname{Xxxx. Xxx. Xxx. Xxx.}
\jvol{AA}
\jyear{YYYY}
\doi{10.1146/((please add article doi))}
\newcommand{\etal}{\textit{et al.}}
\newcommand{\ie}{\textit{i.e.},}
\newcommand{\eg}{\textit{e.g.},}

\newcommand{\revision}[1]{#1}

\begin{document}

\markboth{Bourassa et al.}{Learning the principles of T cell antigen discernment}

\title{Learning the principles of \\ T cell antigen discernment}

\author{François X. P. Bourassa$^1$, Sooraj Achar$^{2,3}$, Grégoire Altan-Bonnet,$^2$ and Paul François$^{4,5}$}

\begin{abstract}
T cells are central to the adaptive immune response, capable of detecting pathogenic antigens while ignoring healthy tissues with remarkable specificity and sensitivity. Quantitatively understanding how T cell receptors (TCRs) discriminate among antigens requires biophysical models and theoretical analysis of signaling networks. Here, we review current theoretical frameworks of antigen recognition in the context of modern experimental and computational advances. Antigen potency spans a continuum and exhibits nonlinear effects within complex mixtures, challenging discrete classification and simple threshold-based models. This complexity motivates the development of models such as adaptive kinetic proofreading, which integrate both activating and inhibitory signals. Advances in high-throughput technologies now generate large-scale, quantitative datasets, enabling the refinement of such models through statistical and machine learning approaches. This convergence of theory, data, and computation promises deeper insights into immune decision-making and opens new avenues for rational immunotherapy design.
\end{abstract}


\maketitle

\affil{$^1$ Joseph Henry Laboratories of Physics, Princeton University, Princeton, NJ, USA, 08544}
\affil{$^2$ Immunodynamics Group, Laboratory of Integrative Cancer Immunology, Center for Cancer Research, National Cancer Institute, Bethesda, MD, USA}
\affil{$^3$ Kennedy Institute of Rheumatology, Nuffield Department of Orthopaedics, Rheumatology and Musculoskeletal Sciences, University of Oxford, Oxford, UK}

\affil{$^4$  D\'epartement de Biochimie et Médecine Mol\'eculaire, Universit\'e de Montr\'eal, Montr\'eal, QC, Canada; email: paul.francois@umontreal.ca}

\affil{$^5$ MILA Québec}

 \newpage

\tableofcontents

\input{Section_0_Introduction}

\input{Section_1_TCRBiophysics}

\input{Section_2_AKPR}

\input{Section_3_ML}

\input{Section_4_Prospectives}



\section*{DISCLOSURE STATEMENT}
The authors are not aware of any affiliations, memberships, funding, or financial holdings that might be perceived as affecting the objectivity of this review. 


%
\bibliographystyle{ar-style6}
\bibliography{References/ARBiophysics}

\end{document}

%% file: Section_0_Introduction.tex
\section*{Introduction.}

The immune system is in charge of defending the organism against external pathogens, or internal threats (such as cancer), while promoting homeostasis and repair for healthy tissues~\cite{pradeu_philosophy_2020}.  Understanding these differential functions remains a challenge, but decades of research have delineated how immune \revision{functions are} highly organized, multilayered, and reactive over time scales ranging from seconds to decades~\cite{mayer_how_2015, mayer_how_2019}. Consequently, one can conceptualize the immune system as a 'liquid brain' \cite{sole_liquid_2019}, and reframe the quest for understanding immunology as a theoretical problem in biological decision-making. In the same way that neurons are the fundamental units of the brain, T lymphocytes are the fundamental units of immune signal processing. T cells are generated via a complex thymic process that randomizes and selects \revision{varied T cell receptors (TCR), distinct} for each individual T cell \cite{wylie_sensitivity_2007, butler_quorum_2013, murugan_statistical_2012, elhanati_predicting_2018}. Within an organism, there is a \revision{large diversity} of TCRs, such that T cells can 'cover' billions of possible immune challenges \cite{de_boer_how_1993, marcou_high-throughput_2018, mora_how_2019, sethna_olga_2019, mason_very_1998}. 

When T cells are activated, an immune response is triggered, possibly leading to an organismal response over days and even years (with the establishment of immune memory)~\cite{mayer_regulation_2019, lam_guide_2024, soerens_functional_2023}. Understanding immune responses starts with understanding the precise computation performed by those first responders~\cite{jobin_distinct_2025}: \revision{returning to} the analogy with neuroscience (and machine learning), can we build and model an immune 'perceptron' that translates molecular inputs into systems-level responses?
 
It is generally assumed that \revision{the} initial T cell computation relates to a fundamental concept of immunology known as self/non-self discrimination \cite{langman_minimal_2000}. TCRs interact with presented peptides (peptide-MHC ligands), and based on those interactions, T cells can be activated  (\revision{by} nonself) or not activated (\revision{by} self). However,  it is becoming increasingly clear that self/non-self classification presents multiple issues~\cite{pradeu_definition_2006} : it is possibly tautological (non-self triggers immune response by definition) and (too) binary -- we will argue that there are indeed many more categories. Recent theoretical works have further suggested that fuzziness is inherent to antigen discrimination, as self and nonself sequences appear statistically \revision{similar}~\cite{mayer_how_2025}, and more dynamical effects might be at play~\cite{pradeu_speed_2013}. For these reasons, in this review, we suggest renaming the self/non-self discrimination capabilities of T cells as ligand \emph{discernment}, to reflect on T cells' ability to make a smart and contextual decision about antigen immunogenicity. Indeed, as detailed later in this review, the notion of antigen ``discrimination'' does not capture the continuous nature of T cells' recognition of peptide-MHC ligands. While classical studies of T cell functions focused on discrete classes of antigen (self vs non-self~\cite{burnet_production_1949}, agonist vs antagonist~\cite{kersh_partially_1999}, positively-selecting vs negatively-selecting ligands~\cite{daniels_thymic_2006}, etc.), we posit that a refined quantitative understanding of TCR signaling will better account for the subtleness of T cell functions.

In the first part of this review, we summarize the experimental facts about T cells' ability to discern between ligands, in particular as it relates to the biophysics of interactions \revision{between the TCR and its ligands}. In the second part, we present theoretical attempts at deriving a phenomenological model to account for all the hallmarks of TCR ligand discernment. In the third part, we review how recent advances in automation and machine learning allow us to better understand the biophysics of TCR activation, opening up new avenues of research in this area. Finally, we will discuss the way forward in developing predictive models and contributing to the field of vaccine and immunotherapy engineering.

%% file: Section_1_TCRBiophysics.tex
\section{The biophysics of ligand discernment by T cells.}

In this section, we review the state of the field of TCR signaling (from structural considerations to cell signaling to biochemistry and first attempts at modeling in biological physics). Our goal is to emphasize how much of the molecular and immunological details are already known and cataloged; yet, it is the mechanism and function of antigen discernment by TCRs that remains mysterious, calling for more quantitative and theoretical understanding.  

\subsection{A few hard facts about antigen discernment by T cells.}

As summarized in~\cite{altan-bonnet_modeling_2005}, there are three quantitative features of T cells one must keep under consideration.

First, T cells are very fast in their ability to respond and to discern between agonist and non-agonist pMHC: experimentalists can detect phosphorylation of the TCR within seconds of pMHC engagement, macroscopic calcium influx and ERK phosphorylation come about within 10 seconds, and release of cytotoxic granules and inflammatory responses are measured within minutes.

Second, T cells are very sensitive, with a single pMHC being sufficient to activate the TCR signaling cascade. Hence, T cells operate at the physical limit of biochemical \revision{detection}, akin to single-photon \revision{counting} in rod cells of the retina, as discussed by Bialek~\cite{bialek_biophysics_2012, rieke_single-photon_1998}. This observation was obtained with careful calibration of radiolabeled ligands in bulk~\cite{sykulev_evidence_1996}, as well as more direct single cell imaging of TCR--pMHC interactions on the surface of cells~\cite{irvine_direct_2002, huppa_tcrpeptidemhc_2010, odonoghue_direct_2013}.

Third, T cells are very specific, with a single mutation potentially converting a self-derived pMHC (that does not trigger immune responses) into a potent agonistic pMHC (that trigger T cells)~\cite{daniels_thymic_2006, zehn_complete_2009, mayer_how_2025}.
Of course, not all single amino acid mutations in the presented peptide alter pMHC potency to such dramatic extent: reduction in experimental costs now allows the systematic testing of all one-amino-acid substitutions in a given agonist peptide~\cite{luksza_neoantigen_2022, kondo_engineering_2025}, and the distribution of antigenic potencies are available for several TCR examples. 

This issue of specificity is critical for applications in the field of cancer immunotherapies, such as the development of bespoke vaccines. In this procedure, clinicians harvest tumors (through biopsies or tumor removal surgery), sequence the tumor transcriptome, model the antigenic landscape and prioritize potentially antigenic mutations (e.g. one mutation of the expressed genome that drives the production of oncogenic protein and the presentation of neoantigen). These mutations are then integrated in mRNA vaccines that are administered to patients to boost anti-tumor T cell response. Recent clinical trials have recorded very promising results in this direction~\cite{sethna_rna_2025, braun_neoantigen_2025}. In the context of this review, we would like to stress how the integration of multiple approaches (from clinical sampling to sequencing, modeling and experimental testing) is required to deliver on the promise of cancer vaccines: at the core, it is the ability of T cells to specifically respond to cancer neoantigens while avoiding self tissues that would deliver the patient-specific eradication of tumors with minimal toxicity. 

At the same time, we want to emphasize how ``hard'' a problem T cells solve when discerning between ligands. Indeed, the sensitivity and speed of TCR triggering are impressive, but its truly distinctive feature is the ability of T cells NOT to respond to non-agonist ligands (e.g. self-derived pMHC) even when those are presented in large quantities; for reference, a dendritic cell can be loaded with more than \revision{$3 \times 10^7$} self ligands and still not elicit any response~\revision{\cite{casasola-lamacchia_human_2021} (50 pmol / million cells; upper bound, during inflammation, MHC class II)}. This distinguishes the TCR from other signal transduction cascades and presents a unique challenge for biophysicists.

\subsection*{Antigen discernment using a dynamic TCR signalosome.}

\begin{figure}[htbp]
\includegraphics[width=5in]{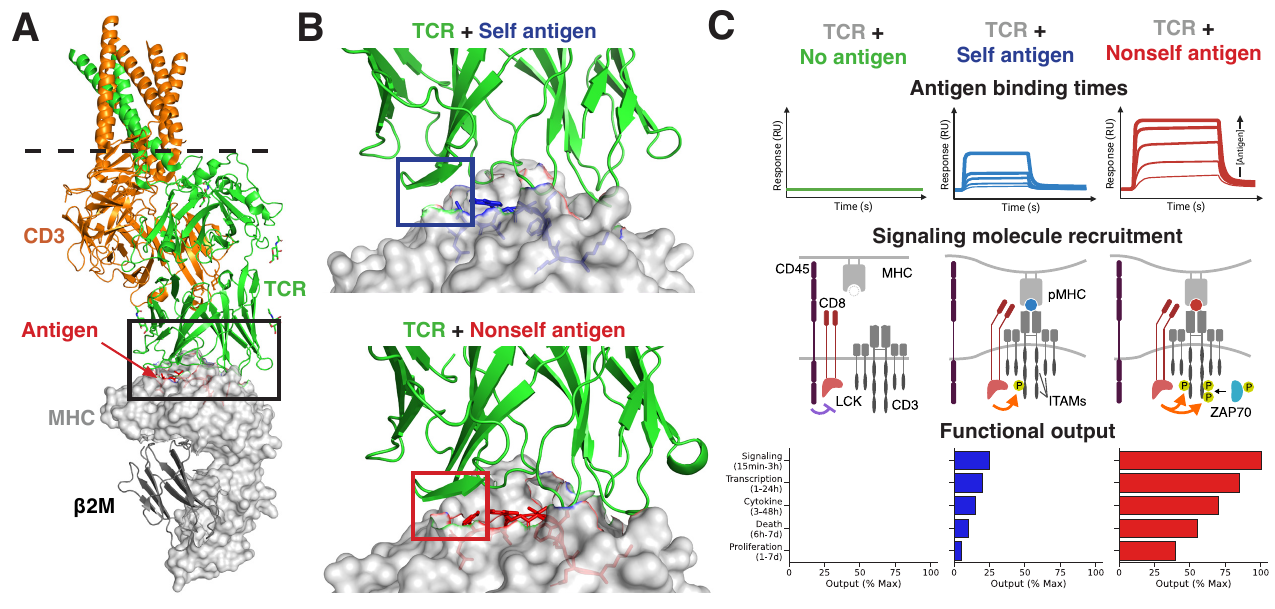}
\caption{\textbf{Overview of TCR structure and activation} (A) Structure of a TCR interacting with Class I pMHC, the initial step for T cell activation (PDB structure 6UK2 from~\cite{devlin_structural_2020}). (B) Close-up and comparison of the presented peptides for anagonist neo-antigen (PDB: 6UK2) and its wild-type counterpart (PDB: 6UK4): note that, despite structural similarities, the neo-antigen pMHC is a 20-fold better binder to TCR compared to the self pMHC, and \revision{200}-fold more potent in activating T cells~\cite[Fig.~2A]{devlin_structural_2020} (C) Sketch of the biophysical/functional characteristics of ligand discernment by TCR. We sketch 3 discrete categories of antigen (no antigen, self antigen, neo-antigen) that do not bind at all, or bind mildly and strongly respectively (top row), inducing increasing degrees of phosphorylation on the TCR and associated proteins (middle row) with different functional outputs (bottom row).
\revision{Note that the graphs in (C) are cartoon illustrations rather than actual data.}    
}
\label{fig:TCRcartoon}
\end{figure}

The TCR signaling machinery comprises a large number of receptors and signaling molecules (from kinases to adapters and scaffolding molecules) that self-assemble when T cells engage a target cell (Figure~\ref{fig:TCRcartoon}). 

The core component is the TCR$\alpha$,TCR$\beta$ heterodimer and its associated chains (two CD3$\epsilon$, CD3$\delta$ and CD3$\gamma$, as  well as two CD3$\zeta$ chains). These associated chains harbor, on their intracellular segments, tandem phosphorylation sites called immunoreceptor tyrosine-based activation motifs (ITAMs), which are short repeated amino acid sequences each containing two phosphorylatable tyrosines. Each $\epsilon$, $\delta$, and $\gamma$ chain contains one ITAM, while each $\zeta$ -- remarkably -- has three, for a total of 10 ITAMs and thus 20 phosphorylatable tyrosines. Models of TCR signaling must accommodate a fundamental aspect of the TCR complex: the TCR itself has no cytoplasmic domains and cannot trigger the signaling cascade on its own; additionally, the TCR signaling complex does not have any signaling capability, as it depends on kinases (e.g. Lck, ZAP70) that get recruited (directly or via CD4/CD8 coreceptors) to phosphorylate ITAMs and to accrue additional adapters (e.g. LAT, Grb2/SOS) for further downstream signaling and gene regulation. 

Classical models of signal transduction in biological physics, such as the bacterial chemotaxis pathway, are relatively ``simple'' in their input/output relationship: the receptor binds one or a few small chemoattractant molecules with high specificity and relays changes in concentration to adjust the tumbling rate of the flagellum motor~\cite{bialek_biophysics_2012}. In contrast, what the TCR signaling complex senses and what it relays in the cell is multifaceted: both quality, quantity and stability of the peptide MHC antigens can be varied widely, and very diverse signaling responses get triggered by the TCR (namely, calcium influx, Erk, p38, JNK and MAPK phosphorylation, and NF$\kappa$B phosphorylation, to name a few). Hence, it is the \revision{complex and combinatorial} nature of the TCR \revision{signalosome} that must spark immunologists' and biophysicists' interest, as it most likely accounts for the phenotypic richness of the signal transduction downstream.

\revision{
The collection of facts about the TCR signalosome presented above provides a detailed molecular description, but it does not, on its own, reveal the principles underlying antigen discernment. Two main classes of explanations have been proposed: structural ones, which rely on conformational changes of the TCR to reflect antigen quality and quantity, and kinetic ones, in which TCR signaling dynamics process these antigenic properties. We now consider each hypothesis in turn. }

\subsection{Is there a structural explanation for ligand discernment?}

There has been a legitimate search for a structural explanation for the outstanding capabilities of the TCR signalosome. Indeed, uncovering a conformational change in the TCR recognition domain associated with agonist pMHC engagement would explain several features of T cell function at once. Because of its molecular nature, \revision{a} conformational change could take place within milliseconds, accounting for the speed of T cell activation; \revision{moreover,} single amino acid mutations in the MHC-presented peptides could be sufficient to drive large (specific) conformational differences in the engaged TCR (as demonstrated in the case of affinity maturation of antibodies against soluble antigens by B cells), and single peptide-MHC could be sufficient to activate the T cell response.

However, structural insights would fail to explain many other experimental facts. First, TCR ligand discernment is highly tunable: the same set of ligands can act as agonists, partial agonists, or non-agonists, depending on the stage of differentiation of T cells; ligands can go from being promiscuous, to sharply specific, to promiscuous again as T cells mature from thymocytes to na\"ive \revision{and} then to memory phenotypes. Second, TCR ligand discernment operates along a continuum of antigenicity, rather than via the digital/all-or-none switch one would expect from a conformational change. Third, conformational changes alone can hardly account for thymic selection, during which the same TCR engages self ligands to drive positive selection, but limits its response to the very same ligands once released to the lymphoid organs and tissues~\cite{daniels_thymic_2006}.  Most stringently, one would need to model how somatically-mutated TCRs would evolutionarily converge to endow T cells with the ligand discernment required by the immune system. 

Structural analysis of the rigidity of the TCR (unbound or bound to peptide-MHC ligands) seems to challenge the notion of peptide-specific conformation changes upon engagement. This rigidity has been pointed out by finding similar conformations in all structures derived from X-ray crystallography (around 120 TCR--pMHC pairs as of today) and from cryogenic electron tomography (which has higher fidelity in preserving structure)~\cite{susac_structure_2022}, (Figure~\ref{fig:TCRcartoon}A). Biophysical measurements by Zhu \& colleagues uncovered the existence of catch bonds for TCR bound to pMHC \ie a universal strengthening of the bond upon binding ~\cite{liu_accumulation_2014,choi_catch_2023}. Three experimental facts challenge the relevance of this catch bond formation. First, one can trigger T cell activation simply by crosslinking TCRs with soluble pMHC multimers without applying any force; one does not expect any catch-bond formation in this context. Second, recent structural studies using cryoEM did not uncover any significant conformational change in the TCR$\alpha \beta$ dimer, again in conditions when T cells are getting activated (Figure~\ref{fig:TCRcartoon}B). Third, biophysicists have been studying artificial systems (using the expression of the TCR components in non-T cells~\cite{James_Biophysical_2012} and/or expression of artificial receptors made out of external domains based on nucleic acids, and intracellular domains derived from the TCR~\cite{James_Biophysical_2012}): catch-bonds do not exist in these artificial systems, yet strong ligand discernment can be achieved. 

Along this line of inquiry and supporting the importance of antigen lifetime rather than conformation in T cell decisions, Tischer \etal{}~\cite{tischer_light-based_2019, britain_progressive_2022} developed an optogenetic system in which blue light intensity increases the effective unbinding rate between a ligand and an engineered receptor without altering the mechanical properties of their bond under load. They observed a strong correlation between T cell activation and the ligand half-life as it varied between $0.5$ and $10$ s -- the physiological range for TCR antigens~\cite{kersh_high-_1998} -- without strong dependence on receptor occupancy. 
Their results favored a model of TCR discernment based on binding lifetime rather than on mechanical sensing, catch bonds or conformational changes. 

Hence, our current appraisal of the field is that catch bond formation for the TCR--pMHC complex may modulate the biophysics of these interactions, but it is unlikely to be a structural answer to the problem of ligand specificity in TCR triggering. As discussed below, a biophysical understanding of TCR ligand discernment based on the kinetics of ligand--receptor complex dissociation (Figure~\ref{fig:TCRcartoon}C, top) offers a more universal, flexible and actionable explanation.

\subsection{The combinatorial complexity of phosphorylation patterns.}
\label{ss:combinatorial_complexity}

Upon reviewing the biochemistry of the TCR signalosome, one is struck by the large number of tyrosines associated with the TCR complex, especially with its ten CD3 ITAMs (Figure~\ref{fig:TCRcartoon}C). Experimentalists have generated mouse models to directly test the functional significance of these tyrosines. 

For example, Vignali and coworkers created 25 genetic models~\cite{holst_scalable_2008} where tyrosines of the TCR--CD3 complex are replaced with phenylalanine (a non-phosphorytable substitute of equivalent hydrophobicity and size). These models displayed an extent of T cell proliferation upon TCR triggering that scaled with the number of tyrosines. More strikingly, mice with fewer tyrosines succumbed to multiorgan autoimmune diseases, highlighting the role of the high number of tyrosines in enforcing immunological tolerance. Hence, \revision{these} T cells \revision{passed} negative selection (\ie{} their signaling against self antigens was \revision{below threshold for apoptosis} in the thymus), yet they also increased their proliferative response to ligands in the periphery (i.e. their signaling against self antigens was above threshold in the periphery)~\cite{Guy_Distinct_2013}. This paradoxical result pointed out to enhanced and defective responses at the same time for TCRs with altered phosphorylation capabilities~\cite{Gaud_regulatory_2018}. 

Earlier attempts at dissecting the role of multiple phosphorylations in the TCR signalosome contributed significantly to our understanding of TCR ligand discernment. Kersh \etal{}~\cite{kersh_fidelity_1998} raised antibodies against distinct phosphorylated peptides covering different potential states of the CD3$\zeta$ chains. These antibodies demonstrated that CD3$\zeta$ undergoes a step-wise set of phosphorylation\revision{s}, with antigens of greater potency reaching new phosphorylated states: the authors concluded that CD3$\zeta$ phosphorylation could establish multiple thresholds in activation that explain the fidelity of ligand discernment. 

Follow-up studies by Aivazian \& Stern~\cite{Aivazian_Phosphorylation_2000} documented how CD3$\zeta$ underwent a reversible association/dissociation from  the cell membrane upon phosphorylation. In this model, before antigen engagement, CD3$\zeta$ burrows in the lipid bilayer, with transient excursions in the cytoplasm. Upon TCR--pMHC binding, Lck gets recruited to the engaged TCR, phosphorylates CD3$\zeta$ during a cytoplasmic excursion, and stabilizes this domain out of the lipid bilayer to further engage downstream signaling molecules (e.g. ZAP70, LAT etc.), (Figure~\ref{fig:TCRcartoon}C, center). This elegant model -- confirmed by more recent cryoEM structural analysis -- would account for the sequential nature of TCR phosphorylation, a key aspect of fidelity in ligand discernment. 

Pitcher \& van Oers~\cite{Pitcher_T-cell_2003} proposed that the combinatorial complexity and sequential nature of phosphorylation in the TCR-CD3 associated chains may engage different downstream signaling pathways. Hence, having such a large number of phosphorylated states may explain how T cells multiplex their response (Figure~\ref{fig:TCRcartoon}C, bottom). This observation could be leveraged to reverse-engineer the key modes of TCR activation encoded, for instance, in T cell cytokine outputs~\cite{achar_universal_2022}. 

Finally, a more recent study by Voisinne \etal{}~\cite{voisinne_kinetic_2022} deployed mass spectrometry to analyze the phosphorylation patterns of the TCR signalosome, when triggered by ligands of varied immunological potency. The authors carefully titrated their TCR ligands to achieve comparable receptor occupancy, and demonstrated that TCR ligands built a hierarchy of phosphorylation states that modulated the stoichiometry of TCR-associated proteins (e.g. ZAP70, LAT, SLP76 etc.) with a critical role for CD6 as a negative regulator that limits the response to weak ligands. Such nested assembly of the TCR signalosome putatively relates to the nested functional responses downstream for T cells (e.g. differentiation, cytokine production and proliferation).

%% file: Section_2_AKPR.tex
\section{Theory of TCR discernment}

\subsection{Life time dogma and kinetic proofreading(s)}
\label{sec:akpr}

At face value, T cell antigen discernment is the high-dimensional problem of matching a space of sequences -- presented peptides -- to function -- commensurate T cell activation. However, the binding of antigens (peptides loaded onto MHCs) to surface TCRs acts as a dimensional bottleneck: for a given TCR, the sequence space is likely projected onto a much reduced space of biochemical parameters characterizing the pMHC-TCR interactions kinetics, such as the on- and off-rates, $k_{\text{on}}, k_{\text{off}}$, the number of peptide ligands $L$ presented, and possibly non-linear effects (e.g. catch-bond dynamics \cite{liu_accumulation_2014,choi_catch_2023}).  

In fact, decades of biophysical measurements have established  that the $k_{\text{off}}$, or more intuitively its inverse, the characteristic time for the pMHC--TCR complex $\tau=k_{\text{off}}^{-1}$, is the best correlate with T cell activation \cite{gascoigne_t-cell_2001,feinerman_quantitative_2008, taylor_dna-based_2017, huhn_3d_2025}. Peptides being indiscriminately presented by antigen presenting cells (APCs), we expect $k_{\text{on}}$ to be limited by diffusion on the APC's plasma membrane, without particular differences between peptides~\cite{altan-bonnet_quantitative_2020}. Hence, differences in the binding constants $K_D = k_{\text{off}}/{k_\text{on}} = (\tau k_{\text{on}})^{-1}$ of peptides are at the level of their binding times $\tau$, with peptides of larger $\tau$s being more antigenic (i.e. triggering better activation) while 'self'-like peptides, which are much more numerous, bind only very transiently or weakly to TCRs.

\subsubsection{McKeithan's original kinetic proofreading model}

The fundamental question of TCR discernment is to account for the ability of T cells to respond to ligands with longer $\tau$s (from pathogens \ie non-self) while remaining quiescent when encountering ligands with shorter $\tau$s (from endogenous tissues, \ie self). The first challenging aspect is that typical non-self ligands have binding times only $10 \times$ to $100 \times$ larger than some of the self ligands~\cite{hogquist_strong_1995, kersh_fidelity_1998, pettmann_discriminatory_2021}, not enough, based on a simple binding equilibrium, to prevent spurious T cell activation by self peptides~\cite{ganti_how_2020}. 
Hence, T cell discrimination first appears as a problem of detection/amplification of weak differences in the stability of the pMHC--TCR complex (quantified by $\tau$).

In 1995, McKeithan~\cite{mckeithan_kinetic_1995} took inspiration from the seminal work of Hopfield and Ninio on the mechanism of kinetic proofreading (KPR) for DNA replication, translation, and tRNA loading~\cite{hopfield_kinetic_1974, ninio_kinetic_1975}. In that process, differences in affinity between correct and incorrect enzyme-substrate pairs are amplified by an extra biochemical validation step, thus lowering the rate of error at the expense of irreversibility and energy consumption. By analogy, McKeithan proposed that the TCR signalosome might be implementing a KPR scheme to discern between ligands of different binding times $\tau=k_{\text{off}}^{-1}$: upon binding, a ligand-receptor complex must undergo a sequence of $N \geq 1$ biochemical modification steps, such as ITAM phosphorylation and signaling molecule recruitment, before reaching a final state able to trigger T cell activation (KPR backbone in Fig.~\ref{fig:theory}A).

 McKeithan's model is a set of ordinary differential equations describing the numbers of receptors on a T cell in each proofreading state, $C_n(t)$, $0 \leq n \leq N$  (stochastic formulations can also be given, see \cite{lalanne_chemodetection_2015, kirby_proofreading_2023} for details; we focus on deterministic versions in this review). Using mass-action kinetics to describe the numbers of TCR-pMHC complexes, when a single antigen type of binding time $\tau$ is present in quantity $L$,
\begin{align}
    \dot C_0 &= \kappa L_\mathrm{free} R_\mathrm{free} - (\phi + \tau^{-1}) C_0  \label{eq:c0} \\
    \dot C_n &= \phi C_{n-1}-(\phi+\tau^{-1}) C_n \quad 1 \leq n < N \label{eq:cndiff}  \\
    \dot C_N &= \phi C_{N-1} - \tau^{-1} C_N  \label{eq:cNdiff}
\end{align}
\noindent where $\phi$ is the phosphorylation rate, $L_\mathrm{free}$ is the number of free ligands (out of $L$ presented ligands in total) and $R_\mathrm{free}$ the number of free receptors (out of $R_\mathrm{tot} \approx 10^5$~\cite{kondo_engineering_2025} in total on a T cell), given respectively by $L - R_\mathrm{b}$ and $R_\mathrm{tot} - R_\mathrm{b}$ with $R_\mathrm{b} = \sum_{n=0}^N C_n$ the total number of bound TCRs (in any proofreading state). Each phosphorylation step is assumed to be irreversible, and the only way a complex gets dephosphorylated is upon dissociation of the ligand from the receptor (which occurs with rate $\tau^{-1}$). At steady-state, we have~\cite{pettmann_discriminatory_2021, gaud_cd3_2023}: 
\begin{align}
    R_\mathrm{b} &= \frac12 \left(R_\mathrm{tot} + L + \frac{1}{\kappa \tau} \right)\left[ 1 -  \sqrt{1 - \frac{ 4R_\mathrm{tot} L}{\left(R_\mathrm{tot} + L + \frac{1}{\kappa \tau} \right)^2}} \,\, \right] \approx  \frac{\kappa R_{\mathrm{tot}} \tau}{\kappa R_{\mathrm{tot}} \tau + 1} L  \label{eq:bound_receptors}  \\
    C_N &= \left(\frac{\phi \tau}{\phi\tau+1}\right)^N R_\mathrm{b} \,\, ,  \label{eq:cN}
\end{align}
\noindent which behaves as $C_N \sim L \tau^N$ in the limit where the forward rate $\phi$ is small, receptors are unsaturated by ligands ($R_\mathrm{tot} \ll L$), and $\kappa R_\mathrm{tot} \tau \gg 1$. This formula highlights the geometric amplification of $\tau$, similarly to the original Hopfield-Ninio scheme, so we expect this process to strongly amplify differences in the binding time $\tau$. Several signaling considerations are consistent with a KPR process being the backbone of T cell recognition, such as the processivity of ITAM phosphorylation~\cite{kersh_fidelity_1998, rohrs_computational_2018} and the sequence of signalosome components recruitment~\cite{ voisinne_kinetic_2022, ganti_how_2020}, as well as the rapid dephosphorylation of ITAMs and Lck after ligand unbinding due to CD45 recruitment~\cite{johnson_supramolecular_2000, davis_kinetic-segregation_2006}. As such, multiple models expanding on and modulating the KPR mechanism have been proposed over the years (see e.g. \cite{lever_phenotypic_2014,lever_architecture_2016, moffett_comparing_2025}).

\subsubsection{Reformulating the 'hard' problem: marginalizing concentration}
The original KPR model was first proposed as a way to lower the equilibrium error rate of enzyme-substrate pairing, $\eta = \frac{C_{\mathrm{correct}}}{C_\mathrm{incorrect}}$, with $N$ proofreading steps giving rise to an effective error rate (at best) of $\eta^N$. It is thus tantalizing to think that KPR for T cell discrimination would serve a similar goal, minimizing the ratio of activated TCRs in the presence of self (binding time $\tau_\text{s}$, ``incorrect'' for activation) vs  non-self ($\tau_\text{ns}$, correct) ligands, i.e. $\eta=\frac{C_N(\tau_\text{s}, \, L)}{C_N(\tau_{ns}, \, L)} \sim \left(\frac{\tau_\text{s}}{\tau_\text{ns}}\right)^N \ll 1$ see e.g. \cite{cui_identifying_2018, kirby_proofreading_2023}. However, there are major differences between DNA replication and ligand discernment by T cells. Most importantly, while the different base pairs in DNA replication have comparable and relatively stable abundances, the quantity of antigens (number per APC, $L$) varies greatly, with self antigens (``incorrect'' ones) being up to $10^5$ times more abundant (1 non-self among $10^5$ pMHCs~\cite{altan-bonnet_modeling_2005, de_boer_how_1993}), amply compensating their 10-100 $\times$ smaller $\tau$. Hence, accounting for $L_\text{ns} \ll L_\text{s}$ in the ratio $\eta=\frac{C_N(\tau_\text{s}, \, L_\text{s})}{C_N(\tau_{ns}, \, L_\text{ns})} \sim \frac{L_\text{ns} \tau_\text{ns}^N}{L_\text{s} \tau_\text{s}^N}$, a pure KPR scheme would give an error rate much closer to $1$. 

These remarks motivate reformulating T cell discernment as a ``hard'' problem. In operational terms, TCR activation should be explained by a model receiving, as an input, ligands with quantities $L_i$ and qualities $\tau_i$, and producing an output $O$; this output should be highly dependent only on the presence of non-self ligands with large $\tau$ while being insensitive to the quantities of these ligands. Mathematically, this problem can be formulated using information ~\cite{cover_elements_2006, bialek_biophysics_2012, lalanne_principles_2013} and decision theories \cite{lalanne_chemodetection_2015}. Focusing for now on the response to a single kind of ligand at a time, we seek a model with a (possibly noisy) input-output function $p(O|\tau)$ which provides high mutual information between the input $\tau$ and some network output $O$ : 
\begin{equation}
    MI(O, \tau) = \int \mathrm{d}\tau p(\tau) \int \mathrm{d}O p(O| \tau) \log_2 \left(\frac{p(O|\tau)}{\int \mathrm{d}\tau' p(O|\tau') p(\tau')}\right) \quad (\mathrm{bits})
    \label{eq:mi}
\end{equation}
while, crucially, the input-output relation is marginalized over ligand quantity,
\begin{equation}
    p(O|\tau) = \int \mathrm{d} l p(O|l, \tau) p_L(l|\tau)
    \label{eq:marginalization}
\end{equation}

\begin{figure}[htbp]
    \centering
    \includegraphics[width=1.3\textwidth]{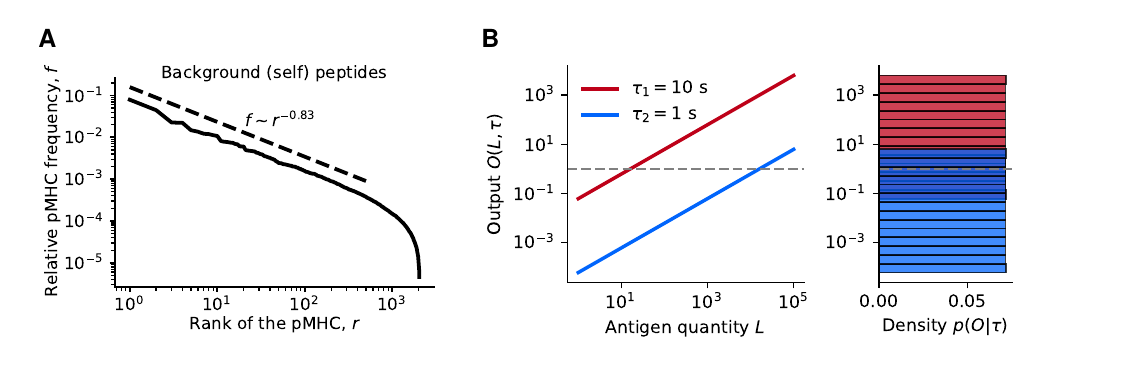}
    \caption{Ligand quantity poses a hard problem for TCR discernment. Adapted from~\cite{bourassa_low-dimensional_2024}.
    \textbf{(A)} Relative frequency distribution of class I peptide MHCs on the surface of antigen presenting
cells (melanoma cells here). Replotting ``background peptides'' (self) data in the ``DMSO'' control condition, averaged over three repeats, from figure 1D of~\cite{stopfer_absolute_2021} (in Dataset\_S02). We notice a power-law behavior for the first $10^3$ peptides, followed by a sharp cutoff. Of note, these abundances are approximate, based on direct spectrometry readouts that may be biased by, e.g., unequal processing efficiencies in the mass spectrometry pipeline for different peptides~\cite{stopfer_multiplexed_2020}. 
\textbf{(B)} Illustration of the antigen quality discrimination problem. (Left) Pure KPR model ($N=4$, $\phi = 0.1$) output curves for two antigen qualities (red: $\tau_\text{s}=10$ s, blue: $\tau_\text{s} = 1$ s) over a wide range of antigen quantities $L$. (Right) Marginalization of the TCR output distribution over the range of $L$, to obtain $p(O|\tau)$. With an absolute discernment model, these distributions would not overlap for different antigens. 
}
    \label{fig:ligand_quantity}
\end{figure}

It is presently unclear what the ``right'' distribution of ligand quantities, $p_L(l|\tau)$, should exactly be. Biophysical evidence indicates that it could generally be a broad, long-tailed, non-Gaussian distribution, such as a log-normal~\cite{feinerman_variability_2008, lalanne_chemodetection_2015} or power-law distribution~\cite{stopfer_absolute_2021, bourassa_low-dimensional_2024} (figure~\ref{fig:ligand_quantity}A). Self peptides being much more abundant, $p_L(l|\tau)$ should be skewed towards lower $\tau$. 

\subsection{Adaptive kinetic proofreading (AKPR) for T cell recognition}

The ``hard'' problem of TCR discernment arises from eq.~\ref{eq:marginalization}, because large variations in antigen quantity create overlap in the distribution of outputs $O$ for a pure KPR scheme where $C_N \sim L \tau^N$ (Fig.~\ref{fig:ligand_quantity}B, right). 
To solve this issue, we are seeking alternative models that are sensitive to few non-self ligands but unresponsive to many self-like ligands \cite{feinerman_quantitative_2008}, a property we refer to as 'absolute discernment'~\cite{francois_case_2016}. Graphically, this property corresponds to a vertical boundary in a phase diagram of immune responses as a function of antigen quality $\tau$ (horizontal axis) and quantity $L$ (vertical axis; a pure KPR model fails to produce such a boundary (Fig.~\ref{fig:theory}B). 

\subsubsection{A first solution: kinetic proofreading with feedback}
\label{subsubsec:2005model}

A first solution to this problem was provided by a seminal theory/experiment work by Altan-Bonnet and Germain, introducing a systems biology model of T cell activation~\cite{altan-bonnet_modeling_2005}.  This detailed biochemical network effectively consists in a kinetic proofreading backbone with two main additions:
\begin{itemize}
    \item TCRs are coupled by the SHP-1 phosphatase, which is activated by an early complex in the kinetic proofreading cascade and dephosphorylates TCR complexes. This implements a negative feedback, following experimental evidence of inhibition and antagonistic interactions mediated by SHP-1~\cite{plas_direct_1996} and partially phosphorylated ITAMs~\cite{sloan-lancaster_partial_1994, reis_e_sousa_partial_1996, madrenas__1995, kersh_partially_1999}; the cross-receptor nature of this interaction was clarified by using (artificial) dual TCR cells \cite{dittel_cross-antagonism_1999,germain_dynamics_1999}
    \item the final step of the proofreading cascade activates a MAPK kinase, eventually responsible for ERK phosphorylation, which self activates and abrogates the SHP-1 inhibitory feedback \cite{stefanova_tcr_2003}.
\end{itemize}

This complex model achieved absolute discernment as a function of antigen quality, predicting a critical binding time (around $\tau_c =3 \, s$) below which even high doses of antigen cannot elicit a response.  The activation threshold can be altered by modulating intrinsic ERK signaling, SHP-1, or CD8 levels, with model predictions matching experimental observations on cell-to-cell variability -- for instance, antigen response is abolished in T cells harboring SHP-1 levels a few-fold above the wild-type average~\cite{feinerman_variability_2008}. 
Hence, it appears that adding feedbacks to KPR can solve the 'hard' problem of TCR discernment independently of ligand quantity. However, the complexity of such biochemical models hinder  their interpretability (a problem also prominent in current machine learning algorithms).

\subsubsection{The adaptive kinetic proofreading framework}
\begin{figure}[htbp]
\includegraphics[width=3.98in]{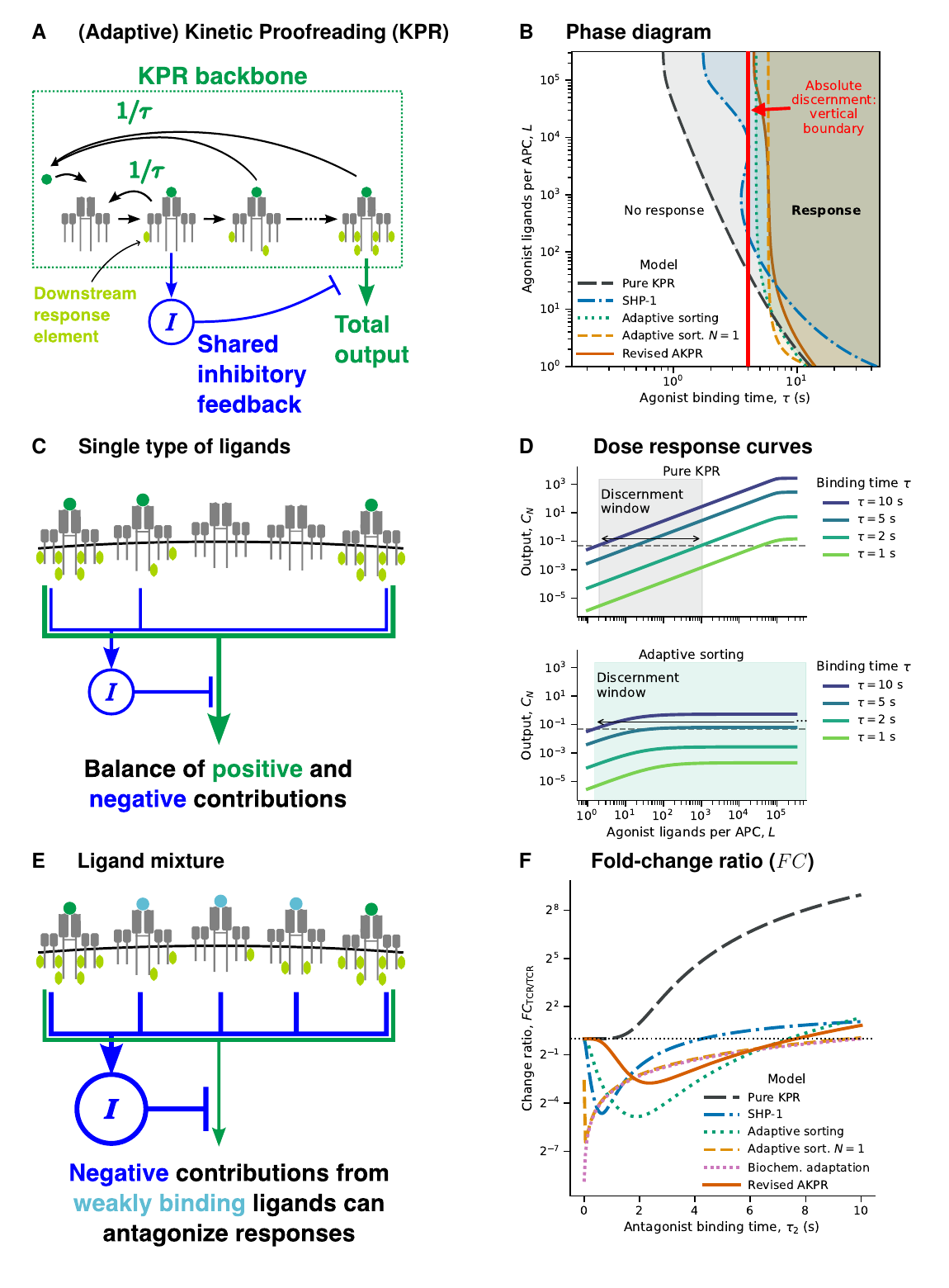}
\caption{Adaptive kinetic proofreading (AKPR) models of T cell ligand discernment. (A) Core reactions in AKPR models, with a KPR backbone and an inhibitory feedback module. (B) Phase diagram of the immune response predicted by different models. The decision boundary is built by setting a threshold for activation on the model output $C_N$, and computing for every antigen quantity $L$ what antigen $\tau$ would be necessary to reach the threshold at that dose. \revision{The ideal boundary for absolute discernment corresponds to a vertical line (\ie{} response above a critical $\tau$, at any $L$)}. (C) Illustration of single antigen responses, where the output is set by the balance of activating and inhibitory signals triggered by the antigen. (D) Model response curves to single antigen types for KPR (top) and adaptive sorting (bottom, $N=6$). The ``discernment window'' is the range of $L$ over which the response is above threshold for agonists ($\tau = 10$ s) but remains below for self-like ligands ($\tau = 2$ s). (E) Illustration of antagonism: weak ligands cannot reach the activated proofreading state, but can trigger the negative feedback. Receptors bound to strong antigens are coupled to this inhibitory module and are thus inhibited, hence lowering the overall cell activation. (F) Fold-change ($FC$) antagonism curves of different models, in the presence of $L_1 = 10$ agonists of $\tau_1 = 10$s, and $L_2 = 5 \times 10^3$ antagonist ligands with $\tau_2$ given by the horizontal axis. }
\label{fig:theory}
\end{figure}

To clarify key mechanisms, Lipniacki \etal{} reduced the model from~\cite{altan-bonnet_modeling_2005} to 37 rate-equations that could reproduce its important properties and capture stochasticity in individual T cells exposed to similar antigens. Yet, the size of their network still obfuscated deeper understanding of ligand discernment. 
In \cite{francois_phenotypic_2013}, a phenomenological model, further simplifying the biochemical network,  was proposed and analytically studied. The model includes a negative feedback with a dephosphorylation reaction $b + \gamma S$ to Eqs. \ref{eq:c0}--\ref{eq:cNdiff}, 
\noindent where $S$ is a pool of active SHP-1 phosphatases shared between receptors, negatively coupling them. The phosphatase is activated by receptors in the first proofreading state, $C_1$. This analytically tractable model recapitulates the predictions of the Altan-Bonnet and Germain model~\cite{altan-bonnet_modeling_2005}, and produces absolute antigen discernment on a wide range of antigen quantities (see Fig.~\ref{fig:theory}C). 

Lalanne and François~\cite{lalanne_principles_2013} built on this insight and leveraged evolutionary algorithms to derive minimal biochemical networks optimizing the mutual information $MI(O, \tau)$ defined in Eq. \ref{eq:mi}. They discovered a variety of KPR-based models with different kinases and phosphatases implementing feedbacks, all with comparable mutual information. In their simplest form, these models boil down to a KPR cascade with the last reaction being mediated by a kinase, $K$, that gets inactivated by an intermediate complex $C_m$, $m < N$:
\begin{align}
    \dot C_N &= \alpha K C_{N-1} - \tau^{-1} C_N  \label{eq:cn_lalanne}   \\
    \dot K &= \epsilon (K_\text{tot} - K) - \delta C_m K  \,\, \Rightarrow \,\, K = \frac{K_\text{tot}}{\delta C_m /\epsilon + 1}  \label{eq:k_adsort}
\end{align}
\noindent In these models, when $L$ is large enough to obtain a strong feedback $K \sim C_m^{-1}$, absolute discernment arises (vertical boundary in Fig.~\ref{fig:theory}B) since $C_N \sim C_{N-1} / C_m \sim  L\tau^N/L\tau^m \sim \tau^{N-m}$. This corresponds to flat dose response curves of the output $C_N(L,\tau)$ as a function of $L$ (Fig.~\ref{fig:theory}D, bottom). Importantly, different antigens (with different $\tau$s) reach distinct plateaus of $C_N$ activation at ligand doses $L \ll R_\mathrm{tot}$ well below the point where receptors get saturated (Fig.~\ref{fig:theory}D).

In these models, there is an incoherent feedforward loop (IFFL) in the KPR cascade \cite{mangan_structure_2003, alon_introduction_2007}, because $C_m$ has both a positive (via the KPR cascade) and negative influence (via $K$) on $C_N$. Such IFFLs are known to be implicated in biochemical adaptation networks~\cite{ma_defining_2009}, and indeed the class of models described by Eqs. \ref{eq:cn_lalanne}-\ref{eq:k_adsort} sort antigens based on quality $\tau$ while adapting (i.e. compensating for) quantity $L$.  Thus, we call these types of augmented KPR models \emph{Adaptive Kinetic Proofreading (AKPR)} models, as schematized in Fig. \ref{fig:theory}A. We notice that the models from \cite{altan-bonnet_modeling_2005, lipniacki_stochastic_2008} appear to perform a similar AKPR computation, as confirmed by automatic coarse-graining methods \cite{proulx-giraldeau_untangling_2017}. Cui and Mehta studied the speed, energy and sensitivity trade-offs for this type of models \cite{cui_identifying_2018}.

\subsection{Antagonism: a 'dog in the manger' effect}
\label{subsec:antagonism}

All AKPR models display an important and biologically relevant property:  ligands below the detection threshold (i.e., not producing a response on their own, so potentially self-like) can hinder activation by ligands above the threshold of activation. Such antagonism in T cell activation is observed experimentally~\cite{hogquist_t_1994, dittel_cross-antagonism_1999,altan-bonnet_modeling_2005}, and is a well-known property of many biochemical detection processes~\cite{reddy_antagonism_2018}.  Torigoe \etal{}~\cite{torigoe_unusual_1998} used the Aesopian fable of the 'dog in the manger' to describe a similar phenomenon in the context of immunoglobulin E receptors (which are related to TCRs).
Like the fable's dog preventing the horse from eating the grain, antagonist ligands actively prevent the agonist ligands to trigger response, while not contributing significantly to the signal, as illustrated in Fig. \ref{fig:theory}D.

Classical KPR models do not exhibit any antagonism (Fig.~\ref{fig:theory}F), pointing out to fundamental limitations of models without feedbacks. It was in fact proven mathematically that antagonism is a necessary byproduct -- an evolutionary ``spandrel'' -- of absolute discernment ~\cite{francois_phenotypic_2016}. Thus, the 'hard' problem of ligand discernment should be solved in the presence of a large amount of self ligands, as is the case in physiological conditions~\cite{ganti_how_2020}.

\subsubsection{Quantifying antagonism with a fold-change ratio}

To quantify antagonism, we define the fold-change ratio for mixtures of two types of ligands (Fig.~\ref{fig:theory}F) :
\begin{equation}
    FC = \frac{\mathrm{Out(mix)}}{\mathrm{Out(Ag)}}
    \label{eq:fc_def}
\end{equation}
\noindent where $\mathrm{Out(Ag)}$ is the response when one type of (agonist) ligand is presented ($\{L_1,\tau_1\}$), while $\mathrm{Out(mix)}$ is the response to a ligand mixture $\{L_1,\tau_1); (L_2,\tau_2)\}$. Antagonism is thus defined by $FC < 1$ (or $\log FC < 0$).

To understand how AKPR models account for TCR antagonism, we consider a generalization of models in \cite{lalanne_principles_2013} (Eqs. \ref{eq:cn_lalanne}-\ref{eq:k_adsort}). The output of AKPR models for a mixture of ligands $\mathcal{C}=\{(L_1,\tau_1); \dots (L_M,\tau_M)\} $ is approximated by:

\begin{equation}
C_N=\frac{\sum_i L_i \tau_i^N}{\sum_i L_i \tau_i^m} \label{eq:cntot}
\end{equation}

\noindent The numerator is the output of a plain KPR cascade, while the denominator quantifies the negative feedback triggered by complex $m$ in the cascade.  For one type of ligand, the response is $C_N=\frac{L\tau^N}{L\tau^m}=\tau^{N-m}$, so it is a pure function of $\tau$ (the quantity dependency cancels out).

For a mixture of two ligands, 
\begin{equation}
    FC =\frac{1}{\tau_1^{N-m}}\frac{ L_1 \tau_1^N+ L_2 \tau_2^N}{L_1 \tau_1^m+ L_2 \tau_2^m} = 1-\frac{L_2\tau_2^m}{L_1\tau_1^m+L_2\tau_2^m}\left(1-\left(\frac{\tau_2}{\tau_1}\right)^{N-m}\right) \label{eq:fc_akpr}
\end{equation}
\noindent \revision{whereby} antagonism occurs \revision{if and only if} $\tau_2<\tau_1$. An interesting particular case is $(m,N)=(0,1)$ : 
\begin{equation}
   FC =  1-\frac{L_2}{L_1+L_2} \left(1-\frac{\tau_2}{\tau_1}\right)
    \label{eq:fc_toy}
\end{equation}
\noindent which can be realized by an (adaptive) ligand-receptor model combined with a $\tau$-dependent degradation of the signaling complex \cite{francois_phenotypic_2016}. Surprisingly, such a model \revision{achieves} absolute discernment without proofreading \revision{in its response to one type of ligand}; however, \revision{for mixtures}, its $FC$ is linear in $\tau_2-\tau_1$, such that the weakest ligands (lowest $\tau_2$) would produce the most antagonism. This dependency would be very problematic in the immune context where $\tau_2 \ll \tau_1$ and $L_1 \ll L_2$: \revision{then, there} would be \emph{no} response to agonists in the presence of self. 
Experimental measurements of antagonism rather show that T cell responses still occur in the presence of very weak self antigens, and that antigens with binding time just below the threshold for activation, $\tau_2 \lesssim \tau_c$, produce the most antagonism~\cite{altan-bonnet_modeling_2005}, thus excluding this regime.

\subsubsection{Proofreading mitigates antagonism and distinguishes AKPR models}

The 'adaptive sorting' model with $N=1$ step initially evolved in \cite{lalanne_principles_2013} for pure antigens displayed too strong antagonism (Fig.~\ref{fig:theory}F). In evolutionary simulations in the presence of self ligands, a solution emerged where the negative feedback in the system is activated by proofreading steps later in the cascade,  and antagonism is considerably mitigated. 

Contrasting with \revision{the $N=1$ model}, self-like ligands are \revision{not} antagonizing the signal \revision{in AKPR models ($FC = 1$ for $\tau_2 = 0$ in Eq.~\ref{eq:fc_akpr})}. Furthermore, \revision{the  curve described by Eq.~\ref{eq:fc_akpr}  is flat close to $\tau_2=0$ as soon as $m > 1$} (Fig.~\ref{fig:theory}F). Hence, to mitigate unavoidable antagonistic effects, $m$ should be large, meaning that the negative feedback should be downstream of a KPR cascade.  

\revision{Moreover, in AKPR models}, there is an intermediate value of $\tau_2$ for which antagonism is maximal \revision{(somewhere between $\tau_2=0$ and $\tau_2 =1 \tau_1$, where $FC=1$)}. For large $m$, this $\tau_2$ is close to antigens at the threshold for activation, $\tau_c$. This was observed qualitatively in the original work of Altan-Bonnet and Germain: the strongest antagonists are also weak agonists\cite{altan-bonnet_modeling_2005}. This hierarchy of antagonism was interpreted in \cite{lalanne_chemodetection_2015} as a way to limit potentially spurious activation by ligands just below the activation threshold. 

All in all, and contrary to what is commonly assumed, this suggests that a high number of KPR steps might not be crucial for absolute discernment (since $N=1$ with feedback can easily realize it, \cite{francois_phenotypic_2016} \revision{and Fig.~\ref{fig:theory}B}). Rather, the most important role of proofreading steps might be to mitigate antagonism by self-like ligands. 

Antagonism also turns out to be important for another, more practical reason: by measuring antagonism in different AKPR models using the FC ratio, we notice that the quantitative details of these curves can vary considerably for different AKPR models (Fig.~\ref{fig:theory}F). To disentangle the precise form of proofreading and feedback modules in models of TCR discernment, precise and extensive measurements of antagonism are thus necessary.

%% file: Section_3_ML.tex
\section{Revisiting antigen discernment in the age of large datasets and artificial intelligence.}
\label{Section_3_ML}

\subsection{Accelerating the study of T cell activation using high-throughput experiments and modeling.}

The previous sections provided a basic mechanistic and theoretical understanding of ligand discernment. Various biochemical models of TCR activation can similarly capture, at a qualitative level, key features like specificity~\cite{mckeithan_kinetic_1995} or antagonism~\cite{francois_phenotypic_2016} (section \ref{sec:akpr}). 

Recent advances with high-throughput experimental tools are providing larger and more quantitative datasets to help falsifying the different models of TCR discernment. High-resolution microscopy can resolve the spatiotemporal kinetics of receptors, ligands, and signalosome formation in the immune synapse~\cite{mcaffee_discrete_2022}. Mass spectroscopy can quantify individual steps of the signaling process in response to different antigens~\cite{voisinne_kinetic_2022}. The potency of these antigens can be measured at very high biophysical precision using optimized surface plasmon resonance protocols~\cite{huhn_3d_2025}, or at high throughput by functionally assessing panels of hundreds of (neo)antigens~\cite{luksza_neoantigen_2022}. Moreover, precisely controlled combinations of ligands can be prepared efficiently~\cite{patel_using_2023} and the fate of individual cells in response to various stimulations can be tracked in vitro~\cite{wither_antigen_2023} and in vivo~\cite{buchholz_disparate_2013, gerlach_heterogeneous_2013}. 

Extensive datasets and machine learning are even more powerful when combined with biophysical theories to pinpoint the mechanisms and design principles of T cell responses~\cite{yuan_kueh_what_2024}. Multiple examples over the years illustrated this approach. For instance, mechanistic model fitting and Bayesian parameter estimation from experimental measurements were crucial in establishing the key roles of signal integration~\cite{gett_cellular_2000, marchingo_antigen_2014} and receptor internalization~\cite{trendel_perfect_2021} in T cell responses to combinations of antigen and costimulatory inputs. Likewise, Wither \etal{}~\cite{wither_antigen_2023} pooled extensive measurements across conditions, machine learning classifiers, and fitting of gene regulation models to disentangle how T cells encode pMHC quality and quantity in their intracellular transcription factor dynamics. Recently, van Dorp \etal{}~\cite{van_dorp_variational_2025} combined a variational autoencoder with a latent Gaussian mixture model and an ODE model of memory T cell compartments to infer subpopulation dynamics from large single-cell datasets collected at different time points. In each case, closely interfaced modeling and data produced immunological insights that could not have emerged from purely theoretical or experimental work. 


\revision{To illustrate the power of allying theory with high-throughput experiments, we now focus on examples from our own research on T cell discernment.} We built the IMMUNOtron robotic platform to streamline the acquisition of large datasets in immunology~\cite{achar_universal_2022}. It is equipped with all attributes experimental immunologists would need on their bench: automatic pipetters, plate handling arm, troughs for reagents, cell incubator, fridge to store samples, centrifuge, etc. (Figure~\ref{fig:Gaud}A). Additionally, we programmed {\it Plateypus}, a Python pipeline to rapidly annotate and to compile large datasets~\cite{achar_plateypus_2021}. This automation improves the multiplexing capability, accuracy and reproducibility of biophysical measurements; we leveraged it to produce new insights into immune cell decision making as detailed below. 

\subsection{Defining and monitoring antigen encoding.}
\label{ssec:antigen_encoding}

Using the IMMUNOtron platform, we introduced the concept of ``antigen encoding'' to describe the global response of T cells to various quantity and quality of antigens~\cite{achar_universal_2022}. We leveraged simple machine learning techniques to compress the high-dimensional bulk cytokine concentration data generated by the IMMUNOtron robotic platform onto a two-dimensional latent space dynamics. We could reconstruct the initial data from the compressed latent representations, proving that our latent space representation is equivalent to the initial data (see Figure~\ref{fig:Gaud}B).

Remarkably, latent space trajectories could be parametrized by a single parameter that increased in proportion to antigenic strength. Applying information theory to this simplified representation, we rigorously defined a continuum of 'antigenicity', divisible into six non-overlapping antigen classes, and thus revealed an antigen 'encoding' in the dynamics of cytokines. This parametrization in a latent space derived from machine learning provided a generative model of cytokine responses along the discovered continuum of antigenicity. This encoding was also ``universal'' to the extent that our model, trained with mouse naïve OT-1 T cells, also applied to human CD8+ T cell blasts and various stimulatory conditions (e.g. different modes of antigen presentation) to effectively rank immune activation according to the quality of the antigens. 
Future work will be necessary to establish how the TCR signalosome explicitly computes such latent variables from antigenic inputs, and how immune cells might themselves biologically decode the information content of cytokines unearthed by our computational analysis.


\subsection{Revisiting the significance of the multiplicity of phosphorylation in the TCR signalosome.}
\label{ssec:multiplicity}

\begin{figure}[htbp]
\includegraphics[width=4.75in]{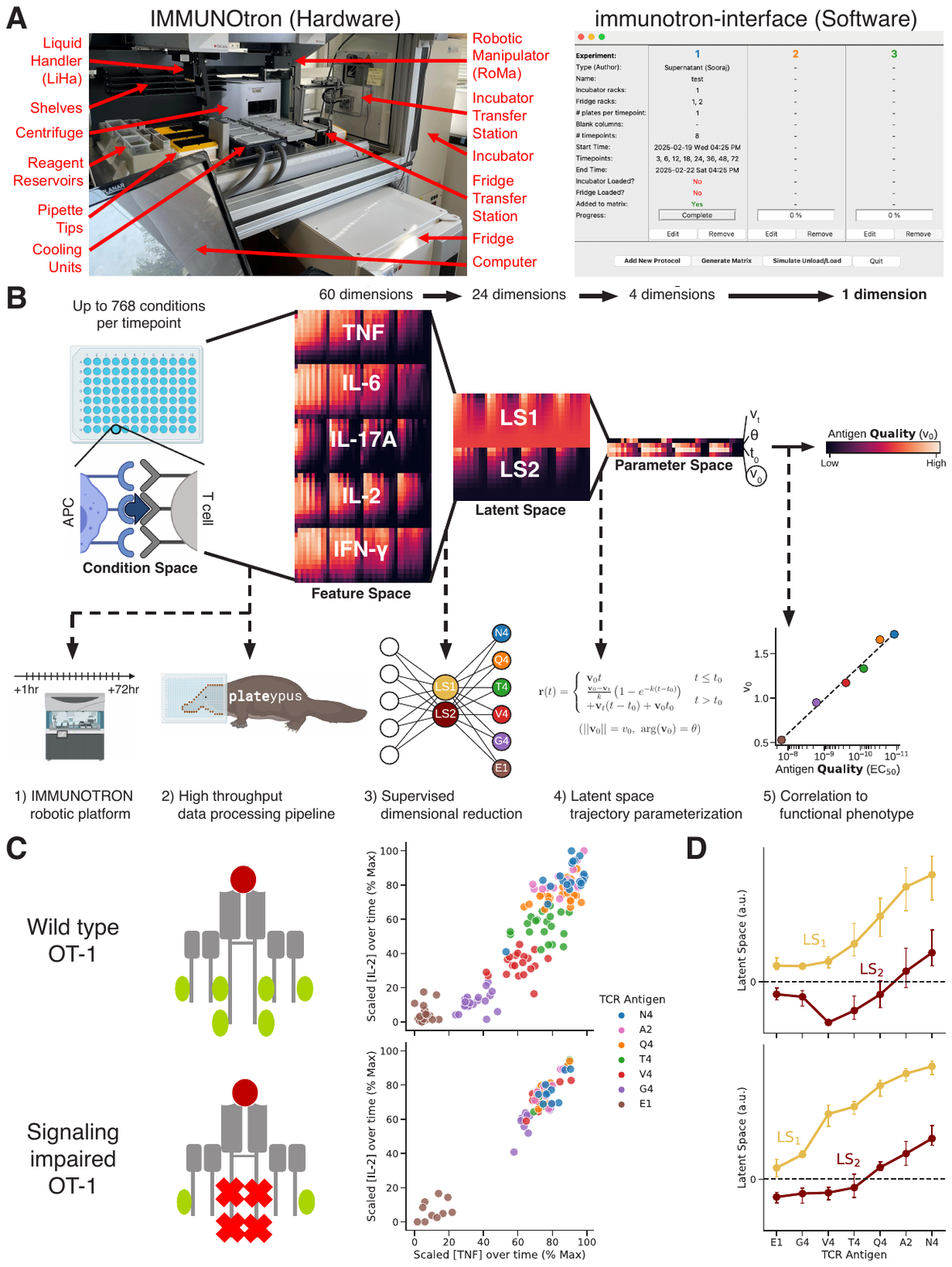}
\caption{\textbf{Applying the IMMUNOtron and antigen encoding framework to analyze ligand discernment in signaling-impaired T cells.} (A) IMMUNOtron robotic platform (left) and software interface (right) used to collect high throughput T cell activation data (B) Schematic of antigen encoding pipeline to deconvolve high throughput T cell activation data (C) Comparison of the cytokine outputs (average secretion of IL-2 vs TNF) for different qualities of ligands (colored) for wild-type (OT1-6Y) (top) and CD3$\zeta$-altered T cells (OT1-6F, bottom): note the loss of resolution for the signaling-impaired 6F TCR. (D) Analysis of the cytokine dynamics using the latent variables reveals a loss of non-monotonic dependency of $LS_2$ with the quality of ligand for the signaling-impaired system.}
\label{fig:Gaud}
\end{figure}

We also applied this IMMUNOtron and machine learning approach (Figure~\ref{fig:Gaud}B) to resolve the paradox of enhanced/decreased signaling of the altered TCR, described in section~\ref{ss:combinatorial_complexity}. Gaud \etal{} re-created a tyrosine--deficient TCR signalosome in the OT-1 transgenic TCR~\cite{gaud_cd3_2023}, by changing the six ITAM tyrosines (Y) of each CD3$\zeta$ chain for phenylalanines (F), thus blunting their phosphorylation potential; this TCR with 4 ITAMs remaining (on the CD3$\delta$, $\epsilon$, $\gamma$ chains) was called OT1-6F, and the wild-type counterpart with 10 ITAMs, OT1-6Y.  We measured the downstream activation of OT1-6F and OT1-6Y T cells against a panel of altered peptide ligands, using the IMMUNOtron platform~\cite{achar_universal_2022} (Figure~\ref{fig:Gaud}B). Strikingly, a simple analysis of integral cytokine secretion demonstrated that OT1-6F responded more strongly to antigens than OT1-6Y, but had very limited discernment of antigen potency (Figure~\ref{fig:Gaud}C). 

We obtained a better understanding of these TCR mutations by compressing the cytokine measurements in the same latent space as in section~\ref{ssec:antigen_encoding} (Figure~\ref{fig:Gaud}D). 
The first dimension (LS1) captured a weighted average of the secreted cytokines, and increased with antigen quality, with stronger response to weak antigens for CD3$\zeta$-deficient receptors (OT1-6F). The second dimension (LS2) had a non-monotonic dependency for wild--type OT1-6Y, reminiscent of a negative interaction akin to inhibitory signals in adaptive kinetic proofreading (AKPR). Strikingly, this relationship was reduced to a simple monotonic function of antigenic strength in the case of the mutant OT1-6F. This change suggested, by analogy with the AKPR networks, that CD3$\zeta$-deficient T cells lacked the negative regulatory loops critical to achieving sharp antigen discernment by T cells.

\subsubsection{From machine learning, back to theoretical modeling.}
We turned this analogy into a direct comparison by probing the biochemistry of these mutated OT1-6F TCRs. We demonstrated the lack of recruitment of the SHP-1 phosphatase upon activation (SHP-1 being a key candidate regulator in AKPR schemes~\cite{altan-bonnet_modeling_2005,francois_phenotypic_2013}). In turn, this defect in negative regulation correlated with enhanced phosphorylation of the CD3 TCR-associated chains, and enhanced recruitment of ZAP70.

We then incorporated these biochemical differences in our phenomenological AKPR model (section~\ref{sec:akpr}) as fewer KPR steps and a blunted SHP-1 activity in OT-6F, accounted for the enhanced sensitivity to weak antigens. A recent model of parallel ITAM phosphorylation can similarly account for these altered single-antigen responses~\cite{morita_parallel_2025}. Importantly, our adjusted SHP-1 model also generated predictions for the response to antigen mixtures; these predictions qualitatively recapitulated the measured patterns measured in the ex vivo stimulation of OT1-6F and OT1-6Y T cells. 
Hence, a machine learning-derived projection of large cytokine datasets pinpointed a key defect in signaling that we further validated theoretically and experimentally. 

\subsection{Unifying principles of antagonism.}

\begin{figure}[htbp]
\includegraphics[width=5in]{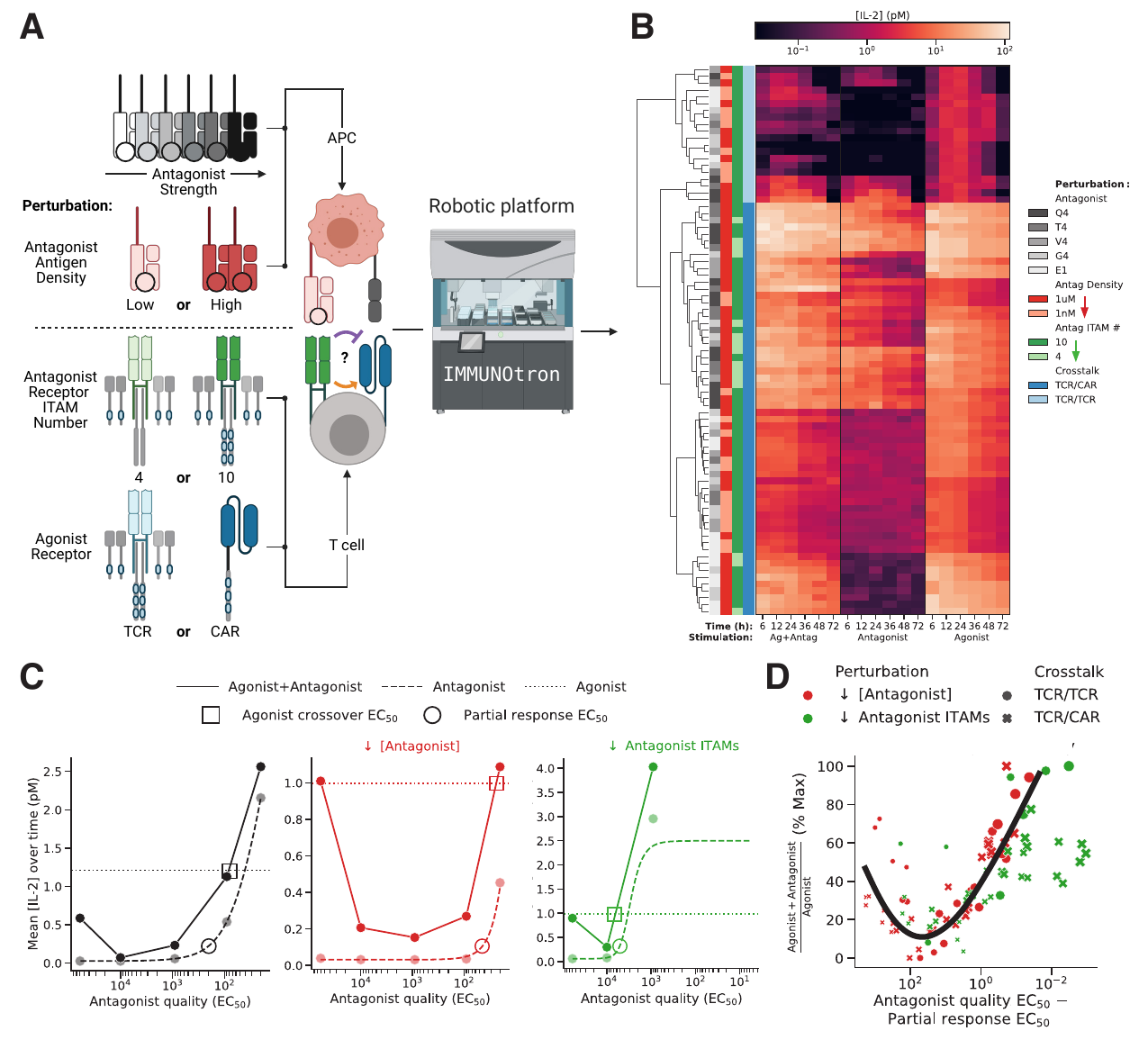}
\caption{\textbf{High-dimensional analysis of TCR antagonism using the IMMUNOtron platform} (A) Schematic of molecular perturbations for CAR T cells applied and analyzed with robotic platform (B) Dataset of all cell activation measurements collected in response to dual TCR/CAR ligand stimulation (C) Combined, antagonist only, and agonist only response curves for various molecular perturbations. ``Agonist crossover EC\textsubscript{50}'': quality of antigen where the agonist response was equal to the ``agonist+antagonist'' response. ``Antagonist partial response EC\textsubscript{50}'': quality of antigen on the antagonist curve where a minimal (10~\% above baseline) response was first achieved. (D) \revision{Scaling the antagonist quality EC\textsubscript{50} by the partial response EC\textsubscript{50} before plotting agonist/antagonist ratios reveals a unified antagonism curve. The solid line is a visual guide. }}
\label{fig:unifiedAntag}
\end{figure}

To further illustrate the power of high-throughput analysis and theoretical modeling for T cell discernment, we present here a re-analysis of our recent study of TCR mediated antagonism~\cite{kondo_engineering_2025}. We used the IMMUNOtron platform to systematically measure cytokine secretion from cultures of mouse OT-1 T cells and OT-1 anti-CD19 CAR T cells, upon activation with a set of altered peptide ligands as well as CD19, both individually and in conjunction. We also tested these T cells in different molecular configurations: by modulating antagonist ligand density or by changing the number of ITAMs on the antagonist receptor (TCR), as detailed in Fig.~\ref{fig:unifiedAntag}A. This resulted in a large matrix of data quantitatively documenting the activation of T cells in response to single or dual ligand stimulations, across a wide range of antigen qualities (Fig.~\ref{fig:unifiedAntag}B).

We then proceeded to plot \revision{against each other the T cell response to the agonist (strong) ligand alone, the antagonist alone, and the mixture of both (antagonism assay), for different molecular settings (Fig.~\ref{fig:unifiedAntag}C). We observed congruent horizontal shifts in the ``antagonist alone'' and``agonist + antagonist'' curves in response to the perturbed conditions. 
Compared to the default condition (left panel), lower concentrations of antagonist ligands (middle panel) require a higher quality of ligand for activation, corresponding to a rightward shift, while reduced TCR ITAM numbers (right panel) have the opposite effect. The mixture curves shift accordingly, as seen from the alignment between the ``partial response EC\textsubscript{50}'' (onset of  activation with antagonists only, circled) with the point where antagonism ends (squared). This coordination is further illustrated by the ``collapse'' of the mixture responses onto a universal curve after rescaling the quality of the antagonist by the partial response EC\textsubscript{50} in each condition (Fig.~\ref{fig:unifiedAntag}D). 
This collapse reveals a strong relationship between the signaling potential of antagonists by themselves, and their ability to antagonize a stronger ligand when presented in a mixture. 
}

These correlations \revision{across perturbations between the antagonism curve and the onset of activation by the antagonist ligands} further reinforce a prediction from Altan-Bonnet and Germain, 2005: the most antagonizing ligands are those closest to "the threshold needed for full signaling"~\cite{altan-bonnet_modeling_2005}. These recent observations opened new avenues for modeling how T cells respond to antigens and, ultimately, for enhanced immunotherapeutic approaches -- in particular, by leveraging the antagonism triggered by self--derived antigens to limit cytotoxic onslaught in healthy tissues~\cite{kondo_engineering_2025}. 

\subsubsection{Return to theory}

Our extensive mapping of TCR/TCR and TCR/CAR crosstalk for T cell activation allowed us to revise previous versions of the AKPR model. We observed that models varied greatly with their predictions on the strength of antagonistic crosstalks, while their predictions of dose response curves were all similar (figure~\ref{fig:theory}B, F). To be specific, the original AKPR models could not account for the (new) experimental observation that a 100-fold decrease in weak TCR antigen levels -- with absolute molecule abundances calibrated experimentally-- still produced significant antagonism, but shifted to stronger antigens as best antagonists~\cite{kondo_engineering_2025}. 

We found that a nonlinear inhibition rate of the output was crucial to match experimental data across the large range of tested antigen quantities, and proposed a biochemical implementation of this mechanism via inhibition of a kinase responsible for proofreading the last reaction step (\cite{kondo_engineering_2025}, STAR Methods). Then, our quantitative measurements of antagonism $FC_{\mathrm{TCR/TCR}}$ and $FC_{\mathrm{TCR/CAR}}$ allowed us to estimate the parameters of this model rigorously via Markov Chain Monte Carlo simulations, whereas phenomenological models from section \ref{sec:akpr} could not quantitatively fit the same data. Hence, combining robot-assisted measurements, machine learning, and biophysical intuition provided a step towards more accurate and predictive modeling. 

%% file: Section_4_Prospectives.tex
\section{Future problems in ligand discernment}

\subsection{Self, the dark matter for T cell ligand discernment}

Most studies in immunology define non-self as immunogenic, \ie as the immunological settings that elicit leukocyte activation. But, as discussed in the review, non-immunogenic ligands (classically the self) still elicit indirect cellular responses by activating negative feedbacks. Hence, the real difference between self and non-self might rather be the balance between overall positive and negative signals, explaining why immunological discernment should be reframed as a continuous problem. The immunological self is not necessary a passive player: it is more like an 'invisible' background, overwhelmingly present, that overall can influence the behavior of T cells in unexpected ways (\eg antagonism). For this reason, calling up an analogy with cosmology, we think of self as an immune 'dark matter': the ubiquitous, unseen yet influential counterpart of immunogenic signals (non-self), whose quantification is crucial to get the full picture of ligand discernment. As with cosmological dark matter, this probing can only be indirect: through antagonistic interactions. To test antagonism in a self background, one should generate various self distributions, and assess their modulation of T cell activation by agonist ligands, which can be modeled and quantified as a fold-change (FC) ratio (Eq \ref{eq:fc_def}). Controlled experiments with automated platforms like IMMUNOtron will prove crucial in this endeavor.

\subsection{Antigen mixtures and vaccine design}

An important and surprising recent finding about T cells' ligand discernement is that immune responses are not binary: a self/nonself dichotomy or a discrete classification of T cell ligands (null, positively-selecting, negatively-selecting) does not properly account for the true spectrum of immune responses. Rather,  antigenicity, T cell population responses ~\cite{achar_universal_2022} and their modulation by antagonistic ligands are continuous features of TCR signaling~\cite{gaud_cd3_2023, kondo_engineering_2025}, which can be ordered in a low-dimensional space. A full understanding of T cell recognition requires a more ``continuous'' mindset: from an experimental standpoint, binary assays to assess antigenicity (such as ELISPOT) are likely too simplified, while from a computational standpoint, future models and machine learning algorithms should predict a spectrum of responses rather than binary outcomes~\cite{visani_t-cell_2025, banerjee_comprehensive_2025}.

Accurately quantifying mixtures of antigens for their immunogenic potential is of crucial importance in vaccine designs, as they may elicit antagonistic interactions. Our own preliminary modeling results suggest that, among peptides one amino acid substitution away from pathogenic agonists and cancer neoantigens, up to 60 \% of these peptides are antagonists~\cite{kondo_engineering_2025}. Weak self-derived antigens can end up being antagonists against the better agonists in the mixture, and this can be tested in dual-stimulation assay. Hence, {\it less is more} when designing bespoke neoantigen vaccines due to the anticipated prevalence of antagonists among single amino acid substitutions. Based on the recent theoretical understanding in TCR ligand discernement, we would caution against administering large mixtures of antigens aimed at maximizing vaccine immunogenicity, because of their potential antagonistic effects.   

\subsection{A theory for adversarial immune strategies}


Going back to theory,  the necessary relation between optimal antigen discernment and antagonism is likely an instance of a more general problem in decision-making. Computer scientists observed that machine learning classifiers 'naively' trained on discrete categories can be easily fooled by very weak but well-designed ``adversarial perturbations''~\cite{szegedy_intriguing_2014}. To solve this issue,  ``adversarial training'' is performed, making sure that classification is still possible in the presence of adversarial perturbations~\cite{goodfellow_explaining_2015}. This strategy is similar to the procedure used in \cite{lalanne_principles_2013} to build networks resistant to antagonism by self-like antigens in TCR signaling. The correspondence between adversarial strategies and antagonism was more rigorously formalized in~\cite{rademaker_attack_2019}. Following inspiration from machine learning, one can design 'adversarial strategies' against AKPR-type models, and demonstrate that there are qualitative changes in the robustness to adversarial attacks in AKPR models depending on the nature of the negative feedback. Such insight resonates with the experimental findings that the complex structure of the TCR signalosome is necessary to achieve enhanced ligand discernment~\cite{holst_scalable_2008, gaud_cd3_2023}.
Notably, the general problem of adversarial examples in machine learning is still unsolved to this date \cite{shayegani_survey_2023, tao_robustness_2024}. It is interesting that the immune system as a whole may have evolved to be relatively robust to 'biological' adversarial attacks -- otherwise, it would be constantly fooled by mutated pathogens with antagonistic properties~\cite{klenerman_cytotoxic_1994}. Seen in this light, the reduction of dimensionality performed by TCR recognition of ligand binding times might be an evolutionary solution to minimize the number of possible adversarial directions in parameter space. 

\subsection{The immune system as a playground for systems biophysics}

For a long time, the dynamical computations performed by the immune system and its underlying logic remained difficult to grasp and to test. In the T cell context, this motivated the design of systems-level models based on first principles~\cite{mckeithan_kinetic_1995} or explicit biochemistry~\cite{altan-bonnet_modeling_2005}.
Within the last few of years, experimental and technological advances (e.g. robotic platforms) have allowed scientists to probe the dynamics of the immune system with unprecedented resolution. Leveraging ideas from machine learning to build data embedding surprisingly revealed an underlying simplicity, with one parameter controlling T cell activation \cite{achar_universal_2022}, consistent with the AKPR families of models~\cite{gaud_cd3_2023}. This alignment between data and simple models in turn allows for more specific tests and improvements of the models (e.g. via antagonism quantification). 

For these reasons, the immune system appears as an ideal playground for systems and theoretical biophysicists, where new principles and frontiers can now be explored theoretically and tested experimentally, and can be extended beyond T cells with similar tools and ideas (e.g. B cell affinity maturation~\cite{merkenschlager_regulated_2025, dewitt_replaying_2025}, regulatory T cells' role in immune tolerance~\cite{wong_local_2021, marsland_tregs_2021}, macrophage polarization to modulate between tissue inflammation and repair~\cite{pizzurro_reframing_2023}). Evolution has sculpted very complex biological networks (signaling and transcription within individual leukocytes, but also at the level of population of leukocytes): it is a gauntlet to biophysicists to generate reductionist/phenomenological models that capture the ``substantifique moelle'' (``substantial marrow'', Rabelais)~\cite[p.~38]{rabelais_gargantua_1998} of the immune system, without missing its key functions. 

Hence, the future looks rich and promising for the field of quantitative immunology: high-throughput quantitative experiments over a wide range of immunological settings will usher testable better theoretical models of immune response, and in turn, such theoretical understanding will generalize better and help optimize immune engineering and control for direct clinical applications~\cite{kondo_engineering_2025}.